\documentstyle[aaspp,12pt]{article}
\def\nfn{$\nu F_\nu$}
\def\ep{$E_{P}$}
\def\dep{$\Delta E_P$}
\def\dpr{$^{\prime\prime}$}
\def\prate{ph-keV$^{-1}$-s$^{-1}$-cm$^{-2}$}

\begin{document}
\title{BATSE Observations of Gamma-Ray Burst Spectra.\\
II. Peak Energy Evolution in Bright, Long Bursts}

\author{L. A. Ford, D. L. Band, J. L. Matteson}
\affil{Center for Astrophysics and Space Sciences 0111,
University of California at San Diego,\\ La Jolla, CA 92093-0111}
\author{M. S. Briggs, G. N. Pendleton, R. D. Preece, W. S. Paciesas}
\affil{University of Alabama at Huntsville, Huntsville, AL 35899}
\author{B. J. Teegarden, D. M. Palmer, B. E. Schaefer, T. L. Cline}
\affil{NASA/Goddard Space Flight Center, Code 661, Greenbelt, MD 20771}
\author{G. J. Fishman, C. Kouveliotou, C. A. Meegan, R. B. Wilson}
\affil{NASA/Marshall Space Flight Center, ES-62, Huntsville, AL 35812}
\and
\author{J. P. Lestrade}
\affil{Mississippi State University, P.O. Box 5167,\\
Mississippi State, MS 39762}

\begin{abstract}
We investigate spectral evolution in 37 bright, long gamma-ray bursts observed
with the BATSE Spectroscopy Detectors. High
resolution spectra are characterized by the energy of the peak of \nfn~and the
evolution of this quantity is examined relative to the emission intensity.
In most cases it is found that this peak energy either rises with or slightly
precedes major intensity increases and softens for the remainder of the pulse.
Inter-pulse emission is generally harder early in the burst. For bursts with
multiple intensity pulses, later spikes tend to be softer than earlier
ones indicating that the energy of the peak of \nfn~is bounded by an envelope
which decays with time. Evidence is found that bursts in which the bulk of the
flux comes well after the event which triggers the instrument tend to show less
peak energy variability and are not as hard as several bursts in which the
emission occurs promptly after the trigger. Several recently proposed burst
models are examined in light of these results and no qualitative conflicts
with the observations presented here are found.
\end{abstract}


\section{Introduction}

Gamma-ray burst continuum spectra provide the most direct information about the
emission processes involved in these violent events. Unfortunately, the
continuum generating process is unknown, making physical interpretation of
spectral observations difficult. A striking feature of gamma-ray bursts is the
temporal variability of spectra both between and within bursts. Therefore, an
empirical study of the dynamics of burst spectra may provide vital clues for
resolving the mystery surrounding these puzzling events.

This is the second in a series of reports describing spectral observations of
gamma-ray bursts as seen by the Burst and Transient Source Experiment (BATSE)
on the {\it Compton Gamma-Ray Observatory} ({\it CGRO}). The goal of this
series is to discover global properties of burst continuum spectra which may
shed light on the gamma-ray burst problem. Gamma-ray bursts remain one of
the least understood phenomena in astrophysics despite years of intense study.
Their heterogenous nature makes classification difficult and few uniform
burst properties are known. Bursts have only been observed above $\gtrsim$2~keV
and no emission has been seen in quiescence (\cite{sch94}). In the first paper
of this series (\cite{band93}, hereafter Paper~I), spectral properties of time
integrated spectra from 54 bright bursts were studied. While spectra are
quite diverse, it was shown that a simple and flexible empirical model
successfully described all spectra. This model is used in the present study to
describe burst spectra and monitor the evolution of the energy at which the
energy flux per logarithmic energy band peaks (\ep, energy of the peak of
\nfn).
In future work, similar evolution studies will be performed with other spectral
characteristics, such as continuum shape or spectral bandwidth parameters.
The data used here are high energy resolution spectra from the Spectroscopy
Detectors (SDs) with moderate time resolution ($\geq 0.128$s). Subsequent work
will use medium energy resolution data with fine time resolution from the
BATSE Large Area Detectors (LADs) to examine weaker bursts and explore
techniques using time-tagged counts from the SDs to improve time resolution.

The study of burst spectral evolution has a long history. Golenetskii et al.
(1983, hereafter G83) examined two-channel data covering $\sim 40$-700 keV
with $\sim$0.5 s time resolution from five bursts observed by the Konus
experiment on {\it Venera} 13 and 14. Spectra were described by an optically
thin thermal bremsstrahlung (OTTB) model, $N_E(E) \propto E^{-1}\exp(-E/kT)$
\prate. A correlation between luminosity and temperature parameter $T$ was
discovered ($L\propto T^\gamma,~\gamma\approx$1.5-1.7), implying spectral
hardness at a given time is related to the intensity of the burst in a simple
way. Laros et al. (1985) performed a similar analysis using five bursts
observed by the {\it Pioneer Venus Orbiter} ($\sim0.1$-2 MeV, $\sim$0.2 s
time resolution) but found no correlation and speculated that the G83 results
were an artifact of the way temperature was inferred.

Norris et al. (1986, hereafter N86) investigated ten bursts seen by instruments
on the {\it Solar Maximum Mission} satellite using hardness ratios (the ratio
of observed flux in two energy bands). The energy bands used in the hardness
ratio were 52-182 keV and 300-350 keV and the time resolution ranged over
0.128-1s, depending on the data available. They found that individual
intensity pulses evolved from hard-to-soft with the hardness peaking before
intensity.  This implies a more complex relationship between hardness and
intensity than proposed by G83.

Recent studies of spectral evolution have tended to follow the lead of G83 or
N86, obtaining similar results. Kargatis et al. (1994, hereafter K94) fit OTTB
and thermal synchrotron models ($N_E(E) \propto \exp[-(E/E_c)^{1/3}]$ \prate)
to sixteen bursts from the SIGNE experiment,
covering $\sim$50-700 keV with 0.5 second resolution.
A luminosity-temperature corelation was found in seven of the bursts
($\gamma\approx$ 1.4-3.0) but two clearly did not show this correlation
(the remainder were questionable) indicating that while many bursts may have a
luminosity-temperature correlation, the trend is not universal. On the other
hand, Band et al. (1992a) analyzed nine bursts observed by the BATSE
SDs ($\sim$25-1000 keV, $\geq$0.128s) using models of the form
$N_E(E) \propto E^\alpha \exp[-(E/kT)]$ \prate, confirming the results of N86
including $T$ leading the intensity when fine time resolution was available.
Bhat et al. (1994) used hardness ratios with BATSE LAD data to study single
pulse bursts which had a fast rise followed by a gradual decay. The energy
bands used for the ratio were $E$=25-100 keV and $E>$100 keV. The time
resolution of the sample was 64 ms. Hard-to-soft spectral evolution was found
with hardness leading the intensity. The time lag between the peak hardness
ratio and peak count rate was found to be directly correlated with the rise
time of the counting rate.

Although the analyses exemplified by G83 and N86 give apparently inconsistent
results, they can be reconciled by considering the time resolution of the data.
The analysis of G83 and K94 had a time resolution of $\sim 0.5$ s while the
resolution of N86 was as small as $\sim 0.128$ s and Bhat et al. (1994) had
64 ms resolution. Therefore, slightly asynchronous behavior
could be masked by the poor time resolution in the G83 and K94 samples. This
conclusion is supported by the results of Band et al. (1992a), who performed
an analysis similar to that of G83 and K94, but found that the spectral
hardness (parameterized by $T$) led the intensity on short timescales.
The need for fine time resolution is emphasized by Kouveliotou et al. (1992),
who examined 22 bursts observed with the LADs and
performed Fourier transforms on
intensity profiles in different energy bands. The hard band was found to lead
the soft by $\sim$0.1 second. In addition, Kouveliotou et al. (1994a) reported
significant variations in the hardness ratio at the 2~ms level in an extremely
intense burst observed with BATSE.

The spectral analyses of G83 and N86 characterize the spectrum in very
different
ways. Both G83 and K94 use spectral models which are assumed to
describe the gamma-ray burst continuum and whose parameters are interpreted as
physically meaningful. Most often, the model chosen
is one which cuts off rapidly at high energies (e.g., OTTB). It was
demonstrated in Paper~I that such models do not reproduce spectra observed by
the BATSE SDs and conflict with the observation of high energy emission by
other experiments (\cite{matz85}; \cite{sch92}; \cite{han94}). Consequently,
these models cannot be physically correct and though the parameters are
certainly physically linked to the emission process, one must realize that the
parameters do not necessarily have their apparent meaning (e.g., the OTTB
cutoff
energy may not be temperature). On the other hand, hardness ratios (such as
used by N86) do not assume any knowledge of the source physics but
do depend on instrumental properties which are accounted for in spectral
fitting. While this characterization tracks changes in the shape and slope of
a spectrum, the numerical values do not have a quantitative physical meaning
and do not lend themselves to additional insights as a characteristic
temperature or spectral shape can. Therefore it is helpful to use an empirical
model which describes the photon spectrum but does not depend
strongly on preconceived notions of spectrum formation in bursts.

In this work, a description of the photon spectrum independent of physical
models is achieved by fitting the empirical spectral model from Paper~I to
BATSE SD data. This model was shown to adequately describe spectra in the BATSE
energy range and incorporates many simple physical models as special cases,
accommodating uncertainty about the actual continuum emission process. Although
the model is phenomenological, physically meaningful parameters can be
derived from it. In this work, \ep~(the energy of the peak of \nfn) is used to
quantify fitted spectra, a parameter which indicates the energy of maximum
radiated power. This allows a study of spectral evolution in gamma-ray bursts
based on a well-defined physical measure of spectral hardness which does not
depend strongly on assumed emission processes.

In this work, the word hardness is used to describe the energy of \ep. Spectra
in which \ep~ is at high energies are called hard and soft spectra have small
\ep. Before discussing the analysis, the SDs are briefly described ($\S$2). In
$\S$3 the analysis techniques used are presented in detail along with several
consistency tests for both the data and methods. The results are presented in
$\S$4 followed by a discussion of their implications for both burst
phenomenology and recently proposed theoretical models ($\S$5). The entire
work is then briefly summarized ($\S$6).

\section{The Instrument}

BATSE is a set of eight detector modules located on the corners of the
{\it CGRO} spacecraft. Each module contains two detectors: an LAD, optimized
for the detection and location of gamma-ray transients, and an SD, optimized
for energy resolution. The SDs were designed for spectral studies and are used
in this work because of their superior energy resolution to the LADs, which
have large area but are thin (20\dpr~diameter by 0.5\dpr~thick) NaI(Tl)
crystals.

The gamma-ray detector for an SD is a 5\dpr~ diameter by 3\dpr~ thick NaI(Tl)
crystal. Because of their thickness, the response to incident gamma-rays is
roughly constant over a large range of energies and viewing angles. The
SDs independently cover two energy decades in the range $\sim$10~keV--100 MeV
(the exact energy range is determined by a commandable gain setting, with
higher gains covering lower energy ranges). The fractional full width at half
maximum energy resolution of the SDs at 662 keV is $\sim$7\%, with a
$\sim E^{-0.4}$ dependence. A 3\dpr~ beryllium window in the front of the
aluminum
case containing the NaI allows the response to extend down to $\sim$5 keV for
face-on bursts. In general however, the lower level discriminator is set so
that the spectrum is cut-off below $\sim$10 keV in the highest gain setting.
An electronic artifact near the low end of spectra (the Spectroscopy detector
Low Energy Distortion [SLED], Band et al. 1992b) exists in several channels
above the instrument's low energy cutoff which can cause problems for spectral
analysis. In this work, only data which are uncontaminated by this artifact are
used. The energy deposited in the crystal is analyzed into 2782 linear channels
which are rebinned on the spacecraft into 256
quasi-logarithmic channels for transmission to Earth. Despite this compression,
the width of most of the compressed channels is significantly smaller than the
detector resolution and little information is lost in this process.

BATSE enters a 4-10 minute long burst mode when the count rate in at least
two LADs exceeds the background count rate by a predetermined significance
threshold (usually 5.5$\sigma$) on one of three timescales: 64 ms, 256 ms, and
1024 ms. Several data types are collected in this mode. The type used in this
study consists of high energy resolution SD spectra with accumulation time
based on a time-to-spill criterion in the LADs (Spectroscopy detector, High
Energy Resolution, Burst, called SHERB). When the total number of
counts in the LADs exceeds a pre-determined threshold, the current SHERB
accumulation ends and the next begins. A total of 192 spectra are recorded from
the SDs associated with the four most brightly illuminated LADs. Half of these
spectra are from the SD associated with the brightest LAD, one-quarter with
the second brightest, and one-eighth each for the other two detectors. The
length of the accumulation is a multiple of 64 ms with a minimum accumulation
of 128 ms. Spectra are accumulated until either all spectra are exhausted or
the burst mode ends. The total time interval covered by each detector is equal:
in the time required to accumulate one spectrum in the third and fourth
brightest detector, four spectra from the brightest and two spectra from the
second brightest detectors are recorded. Additional details about the BATSE
detectors can be found in Fishman et al. (1989) and Horack (1991).

\section{Analysis}

It was shown in Paper~I that the following empirical model is sufficient to
describe burst spectra:
\begin{equation}
N_E(E)~\biggl({{\rm photons}\over\hbox{keV-s-cm}^2}\biggr)=
\cases{A\biggl({E\over{100 {\rm ~keV}}}\biggr)^\alpha
e^{-E/E_0},&$E \le (\alpha-\beta)E_0$\cr
A^\prime\biggl({E\over{100 {\rm ~keV}}}\biggr)^\beta,&$E>(\alpha-\beta)E_0$\cr}
\end{equation}
where $A,~\alpha,~\beta,$ and $E_0$ are fit to observed spectra and $A^\prime$
is chosen so that the function is continuously differentiable everywhere.
The success of this model can be traced to its flexibility.
Included in equation (1) as special cases are a power law, photon exponential,
OTTB, and a broken power law with continuous transition. Since the emission
process is unknown, equation~(1) is a logical choice for determining
physically meaningful parameters. The quantity of interest in this study is
the energy of the peak of \nfn~(\ep) and since $N_E(E)E^2\propto \nu F_\nu$,
\ep$=(2+\alpha)E_0$ (assuming $\beta < -2$).

Burst data were selected from detectors in high gain states, most covering
$\sim$15-1500~keV with about a third extending to higher energies. This choice
was motivated by the results of Paper~I which demonstrated that for most
bursts the spectral slope changes at energies less than 1~MeV and that the
observed signal is greatest around 100~keV (a detector dependent property, not
necessarily true for the incident spectrum).

Bursts which had many high quality spectra were required so that
\ep~could be reliably determined and meaningful trends discovered.
However, most bursts were not sufficiently long or intense to provide enough
spectra to resolve the evolution. To understand the necessary signal strength,
a Monte Carlo simulation of a typical spectrum was performed to investigate
the reliability of the determination of \ep~as a function of the
signal-to-noise ratio (S/N) in the 60-200 keV energy band. The results of this
simulation are given in Table~1 which shows that a large S/N level is required
to reduce the variance in \ep~to reasonable values.

Even for many strong bursts, most individual SHERB spectra do not have large
S/N levels. However, the quality of the signal can be improved by averaging
spectra together weighted by livetime. An explicit demonstration of
this process for burst 2B910807 is given in Figure~1. In this figure,
\ep~as determined from spectra at the finest time resolution available and from
averaged pairs of spectra are plotted together. S/N in the 60-200 keV band
ranged from 7.0-17.0 with a median of 11.7 for the fine time resolution data
and 9.7-23.1 with median 16.3 for the averaged spectra. It is apparent that not
only do the estimated errors in the determination of \ep~get smaller as spectra
are accumulated, but that the fitted value converges to the average of
\ep~determined from the fine resolution spectra.

After selecting candidate bursts, background models were created using a
channel-by-channel fit of a time dependent polynomial to data before and after
the burst. The order of the polynomial varied from burst to burst; typically
the lowest order which gave an acceptable fit was used (never higher than
fourth order). After subtracting the background, consecutive spectra were
averaged together until S/N$\geq$15 in the 60-200 keV energy band. Although
Table~1 shows that this choice will result in a sizable dispersion in \ep, a
larger S/N level would have essentially destroyed the time resolution in most
of the sample. Therefore, some certainty in \ep~was sacrificed for time
resolution. Exceptions to this S/N criterion were made for a few bursts in
which a long period of low activity was bounded by interesting pulse structure.
In those cases, all data in between the intense periods were averaged despite
a low S/N value. An exception was also made for burst 2B910813 in which
the data did not extend below 100~keV. The S/N criterion for this burst was
S/N$\geq$12 in the 100-200 keV band.

A burst was included in the sample if it survived the process outlined above
with at least six spectra. Since most bursts do not have many spectra with
large S/N, only $\sim$4\% of all bursts observed by BATSE and 17~of the~54
bursts from Paper~I were contained in the sample (this sample includes bursts
which occurred after those in Paper~I). The empirical model (eqn.~[1]) was
fitted to data using forward folding deconvolution (\cite{lor89}). This method
determines the incident photon spectrum by folding an assumed photon spectrum
through a model of the detector response and comparing the result to observed
data. The best fit was found using a modified version of the
Levenberg-Marquardt iterative $\chi^2$ minimization algorithm (\cite{bev69},
p. 237; \cite{pre89}, p. 521). The modifications were the use of model
variances instead of data variances in $\chi^2$ (see the appendix for details)
and stopping criteria based on $\chi^2$, $\chi^2$ per degree of freedom in the
fit ($\chi^2_\nu$), and a test for false minima. The detector response model
included the direct component along with scatter off the spacecraft
(\cite{pen89}) and the Earth's atmosphere (\cite{pen92}). The error in \ep~was
determined through standard error propagation using the error in $\alpha$ and
$E_0$ and the correlation between these two parameters (\cite{bev69}, p. 61).

Each spectrum was fit over the entire energy range above the SLED. The data
were not rebinned from the telemetered format so that in most cases the average
number of counts per bin in the high energy channels was less than a few. In
these channels, the Poisson statistical distribution is not well approximated
by a normal distribution and the $\chi^2$ statistic is therefore not strictly
applicable. This is a common difficulty in high energy astrophysics which can
be avoided by binning the data until the Gaussian limit is reached. To
determine the importance of this problem, a set of Monte Carlo simulations was
performed for several types of binning. The results for four different binning
modes in Table~1 show that the fits are not improved by rebinning the
data even if most high energy bins have only a few counts. There are two
reasons for this. First, the data at higher energies are so sparse that when
the Gaussian limit is achieved, the bins are so wide that the upper power law
in
eqn.~(1) cannot be accurately determined. Second, the fit is largely determined
by the signal below a few hundred keV where the number of counts per channel is
largest (Max(C) in Table~1). The Gaussian limit can also be achieved if there
are a large number of background counts, which can be important at low
energies.

These simulations demonstrate that for the SHERB data type used in this sample,
any binning is valid. To ensure that this was the case for real data, fits
using the telemetered binning and eight broad energy bins were made to
spectra from burst 3B930916. The comparison of the values of \ep~is presented
in Figure~2. \ep~is consistent between the two fits and determined with roughly
equal accuracy. Since the choice of data binning is unimportant, the
telemetered channels were not rebinned because the analysis software was
optimized for this scheme.

One difficulty with using high gain detectors (where the spectrum extends to
low energy) was that occasionally the upper spectral index $\beta$ was larger
than $-2$ which implies that \ep~was beyond the upper end of the fit
range, that the spectrum has more gradual curvature than the
empirical model could accomodate, or
that the high energy signal was insufficient to fix $\beta$. For this work,
$\beta$ was constrained to be less than $-2$ for all fits. In bursts for which
this constraint was important (implying a flatter spectrum), it was found that
either $\beta$ tended to shift rapidly from the upper constraint to the lower
($\beta=-5$) or that it remained at the upper constraint ($-2.01$)
consistently. In the former case, the high energy spectrum was probably not
well-determined and should
be ignored when calculating \ep. For the latter, \ep~ should reflect changes in
the shape (i.e. hardness) of the spectrum since $E_0$ will move to higher
energies to compensate for the inability of $\beta$ to adequately describe
this region. No bursts were observed to show convincing evidence for evolution
of $\beta$ from above $-2$ to lower values. Therefore, variations in the
fitted value of \ep ~should be consistent with the true variations in spectral
hardness if quantitatively inaccurate (with systematically smaller values of
\ep). Omitting the upper power law would force \ep~to higher energies as the
model accounts for power at the high end. However, this simpler model may
overestimate \ep, placing it far above the detector's energy range where the
model is not as sensitive to changes in \ep. As a subset of the empirical
four parameter model of equation (1), the model without $\beta$ is not a more
accurate description of the true photon spectrum. Since evolution
of \ep~ is the primary focus of this study, equation (1) with $\beta$
constrained is used to model all spectra since it remains sensitive to
hardness variations over a broader energy range than a model which cuts off at
high energies.

To test whether changes in the fitted values \ep~ were indeed consistent with
real variations in \ep~for cases where $\beta>-2$, detectors at lower gain were
used to determine \ep~ for bursts in which the constraint was important.
Unfortunately, in all but one case the high
energy signal was too weak to reliably fit \ep ~with any time resolution.
Figure~3 shows a comparison of the high and low gain detectors for the
exception (2B910503). It is apparent from this figure that not only do the two
detectors have similar time histories, but that \ep~ determined by
the high gain detector is not far from the value of \ep~ measured by the low
gain detector (for which $\beta < -2$).
Therefore, it is not expected that qualitative statements about the evolution
of \ep~ depend strongly on the accuracy with which $\beta$ is determined.

That the low gain detector measures a consistently larger value of \ep~than the
high gain detector in the previous example suggests the calculated value of
\ep~may depend on gain setting even though \ep~is well defined in all
detectors. To ensure that this was not the case, the value of \ep~from two
detectors at different gains viewing burst 2B921207 were compared ($\beta<-2$
for both detectors). In this case, spectra accumulated by the high gain
detector were averaged so that the time intervals covered by both sets of
spectra were identical. The comparison in Figure~4 shows that the
determination of \ep~was consistent.

The tests outlined above cover the obvious difficulties involved in the
analysis. Although solutions to the S/N and $\beta > -2$ problems are not
ideal, they do allow the analysis to proceed so that useful information about
burst continua can be derived. Also, the inter-detector test indicates
that SD observations are consistent with each other and the results should not
depend on the variable energy range covered by the SDs.

\section{Results}

Of the 862 gamma-ray bursts detected by BATSE until the end of 1993, only the
37
bursts listed in Table~2 met the criteria outlined above. As expected,
equation~(1) fit the data well, $\chi^2_\nu \lesssim 1$ for nearly all spectra.
For these bursts, \ep ~was compared to the count rate (see Figure~5). Two
bursts which clearly demonstrate the general trends observed in the sample as a
whole are 2B921207 and 2B920525. In 2B921207, it is apparent that \ep~softens
over the whole burst except for the increase along with the intensity 8~seconds
after the trigger. 2B920525 is a burst in which two strong intensity pulses
were resolved. \ep ~softens after intial hardening within both spikes and the
second spike is softer than the first.

The plots of Figure~5 were used to characterize types of evolution in \ep.
The classification categories used were motivated by the work of G83 and N86
in which a spectral characteristic (i.e., OTTB temperature and hardness ratio
respectively) was tied to changes in burst intensity. These categories describe
both the tendency for variations in \ep~ to reflect changes in intensity (G83)
and hard-to-soft evolution (N86). The categories are:
\begin{enumerate}
\item An increase in \ep~occurring in proximity to a major intensity increase.
\ep~and intensity need not be morphologically identical and \ep~can lead or
lag the intensity increase by a small amount ($\lesssim$ 1 s).
\item General softening of \ep~in time outside of intensity pulses over the
entire burst. Regions which do not have a spike-like intensity profile are
compared for this category. Spikes are considered to be short time intervals
(less than a few seconds) in which the count rate was much greater than the
rate in nearby intervals. For broader pulses, the first few seconds after the
maximum count rate was achieved are considered part of the pulse, the remainder
was considered inter-pulse emission.
\item Softening of \ep~within intensity pulse structures. \ep~should be
harder early in the intensity pulse for a burst to fall in this category. It
can be measured only if the pulse is temporally resolved.
\item Later intensity pulses have a softer peak in \ep~ than earlier pulses.
This can only be measured if the burst has multiple intensity spikes.
\end{enumerate}
Bursts were judged as showing the property in question, showing an opposite
trend, or showing neither trend ($+$, $-$, and $0$ respectively in Table~2).
To demonstrate how a burst was categorized, consider 2B921207. This burst
satisfies the first category because \ep~hardens as the intensity increases at
0.5~and 8 seconds after the trigger. The interpulse region lies between
2.5~and 8 seconds, where \ep~clearly softens. Both intensity spikes
(0.5-2.5~seconds and 8-10 seconds after trigger) are temporally resolved and
\ep~softens in each. Finally, the peak \ep~ is softer in the later spike than
in the first.

The rating system used here is admittedly simplistic, but is preferable to
schemes which allow for finer distinctions within a class (e.g., rating bursts
on a scale of 1-10, either discretely or continuously) for two reasons.
First, the sample size is small so that the subdivision of bursts into several
discrete subcategories reduces the significance of the classification.
Second, the mediocre time resolution for some time intervals and sizable errors
for some \ep~create problems for a continuous classification scheme.
Therefore, although the criteria outlined above are subjective, they are
the least problematic given the state of the observations.

Individual bursts are detailed in Table~2 and characteristics for the sample
as a whole are given in Table~3. Included in Table~2 are the median values of
\ep~ and the range in \ep~[\dep$\equiv$max(\ep)-min(\ep)] for each burst. Not
all bursts could be classified because of large errors in \ep, unresolved
structure, lack of structure, poor time resolution, etc. Bursts in which
$\beta$ was fit consistently at the upper constraint are noted in Table~2 as
are bursts which were simultaneously observed in the 1-30 MeV range with the
COMPTEL instrument on {\it CGRO} (\cite{han94}). In all four cases, the fitted
$\beta$ was consistent with power law fits to the COMPTEL data in that for
bursts in which COMPTEL reported a power law of index $>-2$, $\beta$ was
consistently fit to the upper constraint. It can be seen from Figure~5 that
there are several instances in which \ep~was poorly fit resulting in a large
error ($\sigma_{E_P}$) for this quantity. Spectra for which
$\sigma_{E_P}\gtrsim$0.5\ep~were ignored when characterizing bursts and
calculating $\Delta E_P$.

Table~2 clearly demonstrates that increases in \ep ~are associated with rises
in intensity and examination of Figure~5 shows that in many cases, the peak in
\ep~leads the intensity peak. A quantitative analysis of \ep~leading intensity
was not performed since Kouveliotou et al. (1992) found the typical lead time
is
shorter then the resolution of this sample. In addition, the time profile
is extremely complex on short timescales so that a detailed study would
best be performed using data with better time resolution. This
data can be obtained from the LADs and the analysis is currently
being performed (Kouveliotou et al. 1994b).
Table~3 shows that many bursts undergo hard-to-soft evolution
both within and outside of intensity spikes as well as between successive
pulses. Softening of successive spikes is explicitly demonstrated in Figure~6,
where the peak in \ep~ for early intensity spikes is compared to the same
quantity in later spikes in multi-spike bursts. It can be seen that late
pulses are more likely to be softer than those which occurred previously,
suggesting that the maximum possible value for \ep~is bounded by an envelope
decaying in time. The trends noted above suggest that spectral hardness leads
the intensity modulated by a decay envelope.

It should be mentioned that there are bursts in the sample which do not share
the traits outlined above. This shows that while typical burst characteristics
may have been identified, they are not universal on the timescale at which this
sample was studied. However, it is not unreasonable to expect that \ep~leading
intensity and the softening trends noted above will be found in most bursts
when observed at finer time resolution.

The general softening trends in \ep~imply that the time at which emission
occurs relative to the beginning of the burst is an important factor in
spectral evolution. This raises an interesting question: assuming that the
maximum values of \ep~ are
bounded by an envelope decaying in time, then if a burst has a precursor
several
seconds before most of the emission occurs, does this envelope begin to decay
when the precursor happens or is the precursor irrelevant with respect to
spectral evolution? To examine this question, the range in
\ep~and the time at which significant emission occurred relative to the BATSE
trigger were compared (Figure~7). Significant emission was defined as emission
capable of producing spectra which met the S/N criterion outlined in $\S$3.
In Figure~7 the horizontal bar indicates the period of significant emission.
It begins at $t_d$, the time delay between the instrument trigger and
the beginning of significant emission, and has length equal to the duration of
significant emission ($\Delta T$). A large value of $t_d$ results when a long
period of time passes between the trigger (a burst precursor) and
the brightest emission. The data in Figure~7 occupy a triangular region
indicating the sample was devoid of bursts with hard, highly evolving spectra
which began emitting significantly well after the BATSE trigger: the few bursts
with large $t_d$ have small values of \ep~and \dep. To determine the importance
of these bursts, the sample was divided into four groups based on the values of
$t_d$ and \dep. Since there are no clear groupings in Figure~7, the divisions
were $t_d=0$ (significant emission at trigger) and \dep =400 keV, which each
divide the sample in half, producing an unbiased division. As shown by Table~4,
bursts with $t_d \ne 0$ do tend to evolve less, but the significance of this
result is small ($\sim 1\sigma$), implying that the apparent
$t_d-$\dep~association may be a statistical aberration which will vanish when
more bursts with large $t_d$ are observed.

For several bursts, the nonzero time delay was short compared to the duration
of significant emission. This suggests that the measure of time delay should be
the ratio of $t_d$ and $\Delta t$. This definition is preferable to the
absolute measure since it incorporates the timescale of the event from the
trigger to the end of significant emission, allowing comparisons based on the
time the burst began its most active stages and the total length of the burst.
Bursts were divided into two groups based on the value of $t_d/\Delta t$.
Table~5 shows the number of bursts in each group for several values of this
ratio as well as the median and extreme values of \dep~ for each
group. The table shows that late emitters tend to show less \ep~ variability
than bursts which emit promptly after the trigger. Although this is
very suggestive, there is sizable overlap between the groups. Considering the
small size of the sample, this result is not too significant and therefore the
possibility of a dependence of \dep~on $t_d$ should be treated
with some caution.

\section{Discussion}

In this work it was shown that for most gamma-ray bursts, the energy of the
peak of \nfn~ is associated with and sometimes leads the intensity of the burst
modulated by an
envelope decaying in time. Indications that bursts whose main emission occurs
at relatively long times after the trigger are generally soft and tend to
evolve less in \ep~ were also found. In this section the phenomenological
implications of these results are discussed. Some recently proposed burst
models are also examined in light of the observations reported here and no
qualitative conflicts between the models considered and observations are found.

\subsection{Phenomenological Implications}

While it has been known for some time that burst spectra tend to evolve from
hard to soft (N86), the observations presented here show that the time of
emission relative to the beginning of the burst may also determine how the
spectral hardness evolves. For optically thin burst models (in which particles
radiate on timescales much shorter than the resolution of SHERB spectra),
these results imply that as the burst transpires, the emission region
`remembers' previous emission suggesting that either emission comes from only
one region or that different emission regions must be
physically connected. The softening of successive intensity pulses could be
caused by significant change in the emitting region during intense episodes
(such as an expansion or an increase in the number density of particles). The
softness of delayed emission implies that the emission region can evolve
without observable radiation. For optically thick models, a burst could be the
result of a single energy input since the radiation does not escape the region
on short timescales. If the region cools by adiabatic expansion and the surface
of last scattering does not propagate inward to higher temperature regions too
quickly, then observable
radiation would be expected to evolve from hard-to-soft. The softness of
delayed emission could also be explained in this way if the region was not able
to emit significantly during the initial stages of expansion.

The hard-to-soft spectral evolution within pulse structures indicates that
intensity spikes are not symmetric in time. Nemiroff et al. (1994b) reached a
similar conclusion by examining the time structure of intensity profiles. These
observations clearly show that time symmetric models of spectra formation as
well as overall intensity profiles cannot be correct. An example of the kind of
model eliminated by these observations is symmetric beam models in which the
beam sweeps by much faster than the timescale over which the beam changes.

Another implication of this work relates to burst duration classes. The
distribution of gamma-ray burst durations observed by BATSE is bimodal and
short bursts (duration~$\lesssim 2$~s) tend to have larger hardness
ratios than longer ones (\cite{kou93}; \cite{lam94}) where the hardness ratios
were averaged over the whole burst. Based on the large values of \dep~reported
here and the fact that the hardest emission tends to come early in a burst,
short bursts are expected to be harder than long ones since they do not have
softer tails, not necessarily because of an inherent
difference in the radiation mechanism. An example of the effect of time
averaging is burst 2B920525. If spectra from 4.16-6.72~s (the first spike) are
used \ep=650$\pm$70~keV but falls to 421$\pm$25~keV if the entire burst is
integrated into one spectrum. As Norris et al. (1994) also suggested, if
hardnesses are measured over the same absolute time interval (e.g., compare
short bursts to the first few seconds of long bursts), the hardness
distribution may be the same for both classes.

In Paper~I it was shown that \ep~varied considerably from burst to burst
indicating that there is no universal characteristic energy of burst emission.
One conclusion which was drawn from this was that pair processes could be
directly observed only if burst sources had a broad range of redshifts. Such
processes would produce spectra with an annihilation feature at 511~keV and
would cut off above this energy because of pair opacity. The observed
energy of these features depends on the redshift of the source region.
The argument against observable pair processes can be extended by noting the
large changes in \dep~observed within many bursts in this sample (Table~3).
Unless the emission process involves large and rapid changes in redshift, pair
processes cannot be directly observed in burst spectra.

The data in Figure~7 include several bursts in which the emission was prompt
but had small \dep. An interesting possibility is that these prompt emitting,
relatively non-evolving bursts were preceded by a precursor too faint
to trigger BATSE. In this case, bursts which occupy the lower left-hand corner
of the figure are late emitters for which BATSE did not trigger on the initial
event. A less exciting possibility is that bursts which emit late
but evolve considerably exist and there is no relation between emission time
and \dep. Although this type of behavior is not seen, it cannot be ruled out
because of the small sample size and rarity of late emitters (about one strong
burst per year with $t_d>$60s). Therefore, this may remain an open
question even if BATSE detects many bursts over the next few years.

\subsection{Implications for Burst Models}

A plethora of gamma-ray burst scenarios have been proposed (\cite{nem94a}).
However, most attempt to explain only general spectral characteristics and
energetics with little or no mention of how spectra are expected to evolve.
Therefore, several models are examined to find those which can accommodate the
observations presented here. Because BATSE
observations of burst isotropy and inhomogeneity have cast serious doubt on
older models, only those proposed after 1991 (the year BATSE was launched) are
considered. Discussion is further restricted to models which purport to
explain all or most gamma-ray bursts. Although arguments have been made for
two population distributions (\cite{lin92}; \cite{smi93}), the
inability to separate bursts into two uncontroversial classes makes such
discussion premature. Jet models are also avoided (c.f. \cite{bri92};
\cite{der94}) since the phenomenon is poorly understood and the models are
incomplete. `Exotic' sources such as cosmic strings (\cite{pac88}),
strange stars (\cite{hae91}), or primordial black holes (\cite{cli92}) are
also omitted since the existence of such sources is highly speculative.

The distribution of bursts seen by BATSE has forced bursts out to extended
halo or cosmological distances if they arise from a single source population.
Since source regions are inferred to be small from the fast rise
times of bursts, the energy density at the source must be enormous,
particularly if the burst radiates isotropically. Therefore, many models
incorporate optically thick relativistic electron-positron fireballs. As
originally proposed (\cite{good86}; \cite{pac86}), fireballs cannot explain
the observed nonthermal emission and will not be relativistic if the fireball
is contaminated by baryons (\cite{cav78}; \cite{pac90}).

Many different  scenarios have been proposed recently (c.f. \cite{nar92};
\cite{usov92}; \cite{woo92}) which can essentially be reduced to
a fireball which must either form in a region free of baryons or break through
a cloud of baryons to be observed. In these cases, the observed nonthermal
spectra are produced either by several thermal blackbodies at different
temperatures (corresponding to different fireballs) or through interactions
with a strong or turbulent magnetic field. If a number of fireballs occur, then
hard-to-soft spectral evolution could be explained by the creation of low
temperature fireballs or rapid cooling as the burst transpires. For magnetic
interaction models, the general softening of \ep~may reflect a gradual decline
in available energy as the field evolves into a stable configuration.

M\'esz\'aros \& Rees (1992) have proposed a model involving a baryon
contaminated fireball from any source, galactic or cosmological. The baryons in
the fireball interact
with the interstellar medium (ISM) to form a relativistic shock where
gamma-rays are radiated. Hard-to-soft evolution could be explained either by
the most energetic baryons reaching the shock first or the
gradual decline of average energy as more particles encounter the shock. The
softness of late emission could be a consequence of the baryon front losing
energy to the ISM before the shock is fully formed or could arise from the
shock losing energy as it propagates through the ISM.

In a different type of model suggested by Melia \& Fatuzzo (1992), a gamma-ray
burst is generated by sheared Alfv\'en waves from a radio pulsar. The
waves are created by crustal disturbances at the polar cap which flood the
magnetosphere with charged particles. These charges upscatter ambient radio
photons to gamma-ray energies which are beamed outward. Hard-to-soft evolution
may result from the most energetic charges scattering first and successive
spikes soften if subsequent disturbances are less energetic than the
original disturbance. Since the emission is beamed, the hard emission might
occur when the beam was not pointed in our direction. Delayed emission might
be soft because the early hard emission was not beamed in our direction.

\section{Summary}

In this work, the evolution of the energy of the maximum flux per logarithmic
energy band (\ep) in long, bright gamma-ray bursts was studied using high
energy resolution spectra from the BATSE SDs. It was found that \ep~is
coupled with the intensity of a burst (either leading or accompanying intensity
increases) but is apparently modulated by an envelope decaying in time.
These results are consistent with previous studies by N86 and Bhat et al.
(1994). Indications were also found that bursts in which significant emission
is substantially delayed relative to the instrument trigger tend to be soft and
evolve less in \ep~although the significance of this result is low. Several
burst models were examined in light of these
results and it was found that none of those discussed could be ruled out on
the basis of these results alone. However, as the models mature to the point of
more precise spectral predictions, the observations presented here may prove
to be a useful constraint or suggest directions for further effort.

\acknowledgments

We thank D. Gruber, J. Higdon, R. Lingenfelter, and R. Rothschild for helpful
advice and useful information. D. Marsden and D. Potter provided additional
assistance. BATSE work at UCSD is supported by NASA contract NAS8-36081.

\appendix
\section{Model Weights for $\chi^2$}

The Levenberg-Marquardt algorithm (Bevington 1969, p. 237; Press et al. 1989,
p. 521) is a prescription to efficiently minimize
goodness-of-fit statistics. The most common statistic used is
\begin{equation}
\chi^2=\sum^N_{i=1}\biggl[{y_i-y(x_i;{\rm\bf a})\over\sigma_i}\biggr]^2,
\end{equation}
where: $x$ is the independent variable (here energy); $y$ is an observed
variable (count rate in this case); $y(x_i;{\rm\bf a})$ is the value of
the model at $x_i$ for parameters {\bf a}; $N$ is the number of measurements
(here, channels in a spectrum); {\bf a} represents the parameters to be fit;
and $\sigma_i$ are the errors associated with $y_i$. $\chi^2$ is minimized
using an iterative gradient procedure for which the first and second partial
derivatives of $\chi^2$ with respect to {\bf a} are needed. In its usual
implementation, the algorithm assumes that the $\sigma_i$ are taken
directly from the data and are constant.

The problem with using data variances ($\sigma_i$) in equation (A1) is that
$\sigma_i$ weights downward fluctuations in the data too strongly, a result of
the Poisson nature of count data in which $\sigma_i^2\propto n_i$, the number
of observed counts. In this case, a small number of counts create a smaller
than average $\sigma_i$ which causes the associated data points to be more
important in the calculation of $\chi^2$.
Wheaton et al. (1994, hereafter W94) examined this problem in detail using a
Monte Carlo simulation of 24,000 one-parameter data sets and found
that the mean of fitted values deviated from the true value by
$-170\sigma$ (note the downward shift). The solution to this
problem is to use $\sigma_i$ generated from the model under consideration
rather than the data, replacing $\sigma_i^2\propto n_i$ with
$\sigma_i^2(x_i;{\rm\bf a})\propto n_i(x_i;{\rm\bf a})$ (model variance).
When W94 performed the simulation with this new scheme, the bias was
$-0.34\sigma$.

Although W94 demonstrated the importance of using weights derived from
the model, they do not mention a necessary change in the standard
Levenberg-Marquardt prescription to accommodate the new weighting. The
algorithm assumes that $\sigma_i$ is taken from the data so that changes in
the $\chi^2$ surface are contained in only the $y(x_i;{\bf a})$ term in
equation (A1). However, by making $\sigma_i$ model dependent without changing
the partial derivatives, assumed changes in the $\chi^2$ surface caused by
variation in {\bf a} will be incorrect, resulting in slightly off-course
iterations and an incorrect minimum since $\partial \chi^2/\partial${\bf a}=0
can be at the wrong value of {\bf a}. It was found that when this effect is not
accounted for, the true minimum of the $\chi^2$ surface is not reached and the
values of the fitted parameters differ from the true best fit by roughly the
quoted error in the fitted parameters though it is not biased toward higher or
lower values making it difficult to detect in large collections of fits.

Alterations in the second derivative terms only affect
the route used to find the minimum of $\chi^2$ and therefore need not
be corrected for the new weighting scheme (\cite{pre89}, p. 523). On the
other hand, the first partial derivatives need to reflect any changes which
might occur in the $\chi^2$
surface and must be corrected. The corrected version is
\begin{equation}
{\partial\chi^2\over\partial a_j} = -2\sum^N_{i=1}\biggl[{y_i-
y_i(x_i;{\rm\bf a})\over\sigma_i^2(x_i;{\rm\bf a})}\biggr]
\biggl({\partial y_i(x_i;{\rm\bf a})\over\partial a_j}-{[y_i-
y_i(x_i;{\rm\bf a})]\over\sigma_i(x_i;{\rm\bf a})}{\partial
\sigma_i(x_i;{\rm\bf a})\over\partial a_j}\biggl).
\end{equation}
The first partial derivative term in this equation is equal to the partial
derivative specified by the Levenberg-Marquardt algorithm with model variances
replacing data variances but the second term is new. For this work, equation
(A2) can be expressed in a simpler form since background subtracted count
rates are used ($y_i=n_i/f_i-b_i$ where $f_i$ is a term which converts counts
to a count rate and $b_i$ is the background counting rate). In this case,
$\sigma_i^2(x_i;{\rm\bf a})$ becomes
\begin{equation}
\sigma_i^2(x_i;{\rm\bf a})={y_i(x_i;{\rm\bf a})\over f_i}+\sigma^2_{b_i},
\end{equation}
where $\sigma^2_{b_i}$ is the error in the determination of the background rate
in channel $i$. Differentiating (A3) with respect to {\bf a} and substituting
into equation (A2), it is found that
\begin{equation}
{\partial\chi^2\over\partial a_j} = -\sum^N_{i=1}[y_i-y_i(x_i;{\rm\bf a})]
{\sigma_i^2+\sigma_i^2(x_i;{\rm\bf a})\over\sigma_i^4(x_i;{\rm\bf a})}
{\partial y_i(x_i;{\rm\bf a})\over\partial a_j}.
\end{equation}
This expression is equal to the uncorrected expression when
$\sigma_i(x_i;{\rm\bf a})$ is replaced by $\sigma_i$ which allows the standard
algorithms (e.g. Bevington 1969, p. 237; Press et al. 1989, p. 521) to be
modified easily. When this expression is substituted into the algorithm, the
true minimum of the $\chi^2$ surface with model variances is found.

\clearpage

\begin{planotable}{rcccccc}
\tablewidth{425pt}
\tablecaption{Monte Carlo Simulations}
\tablehead{
\colhead{Binning\tablenotemark{1}} &
\colhead{$t$(s)\tablenotemark{2}} &
\colhead{Med(C)\tablenotemark{3}} &
\colhead{Max(C)\tablenotemark{4}} &
\colhead{Med(\ep)\tablenotemark{5}} &
\colhead{Var(\ep)\tablenotemark{6}} &
\colhead{S/N\tablenotemark{7}}}
\startdata
nobin & 0.1 & 0.23 & 4.66 & 259 & 286 & 10.0 \nl
& 0.25 & 0.99 & 7.78 & 285 & 77 & 15.6 \nl
& 0.5 & 2.11 & 16.48 & 265 & 50 & 21.6 \nl
& 1.0 & 4.41 & 30.31 & 251 & 36 & 29.6 \nl
& 2.0 & 9.70 & 43.72 & 250 & 21 & 39.5 \nl
\tableline
half & 0.1 & 1.18 & 9.52 & 271 & 176 & 10.0 \nl
& 0.25 & 2.78 & 26.85 & 259 & 85 & 15.6 \nl
& 0.5 & 4.66 & 45.71 & 256 & 56 & 21.6 \nl
& 1.0 & 9.05 & 86.85 & 252 & 39 & 29.6 \nl
& 2.0 & 21.50 & 219.33 & 251 & 22 & 39.5 \nl
\tableline
full & 0.1 & 2.46 & 15.47 & 252 & 153 & 10.0 \nl
& 0.25 & 4.95 & 36.03 & 244 & 68 & 15.6 \nl
& 0.5 & 8.55 & 72.05 & 253 & 50 & 21.6 \nl
& 1.0 & 19.30 & 141.77 & 251 & 35 & 29.6 \nl
& 2.0 & 45.47 & 321.04 & 250 & 21 & 39.5 \nl
\tableline
eight & 0.1 & 3.29 & 10.52 & 220 & 86 & 10.0 \nl
& 0.25 & 8.54 & 24.35 & 240 & 63 & 15.6 \nl
& 0.5 & 17.67 & 46.76 & 247 & 46 & 21.6 \nl
& 1.0 & 34.85 & 102.28 & 250 & 40 & 29.6 \nl
& 2.0 & 73.27 & 195.23 & 248 & 25 & 39.5 \nl

\tablenotetext{1}{The binning scheme used for spectra:
\begin{verse}
nobin --- Data in native format.\\
half --- Binned to one half of the detector energy resolution.\\
full --- Binned to the detector energy resolution.\\
eight --- Eight broad energy bins.
\end{verse}}
\tablenotetext{2}{The livetime used to create spectra.}
\tablenotetext{3}{Median number of counts per channel in the binned spectrum.}
\tablenotetext{4}{Maximum number of counts per channel in the binned spectrum.}
\tablenotetext{5}{Median value of \ep~(keV).}
\tablenotetext{6}{Variance of \ep~about the median (keV).}
\tablenotetext{7}{Median S/N in the 60-200 keV energy band.}

\tablecomments{200 spectra were created using the model spectrum of eqn.~(1)
with $A$=0.1 \prate, $\alpha=-1$, $\beta=-3$, $E_0=250$ keV (\ep=250 keV).
Non-integral counts result from subtraction of fractional background counts.}

\end{planotable}

\clearpage

\begin{planotable}{ccccccc}
\tablewidth{300pt}
\tablecaption{Classification of the Sample}
\tablehead{
& \multicolumn{4}{c}{Characteristic} & Med(\ep) & \dep \\
\colhead{Burst} &
\colhead{1} &
\colhead{2} &
\colhead{3} &
\colhead{4} &
\colhead{(keV)} &
\colhead{(keV)}}
\startdata
2B910503$^{*\dagger}$ & $+$ & $+$ & $+$ & $+$ & 540 & 1350 \nl
2B910601$^\dagger$ & $+$ & $0$ & \nodata & \nodata & 700 & 950\nl
2B910807 & $+$ & $0$ & $+$ & $0$ & 170 & 175 \nl
2B910814 & \nodata & \nodata & \nodata & \nodata & 490 & 900 \nl
2B910814C$^\dagger$ & \nodata & $+$ & \nodata & \nodata & 1400 & 1300 \nl
2B911031$^*$ & \nodata & $+$ & \nodata & \nodata & 400 & 850 \nl
2B911106 & \nodata & $0$ & \nodata & \nodata & 165 & 120 \nl
2B911118$^\dagger$ & $+$ & $+$ & $+$ & $+$ & 180 & 440 \nl
2B911126$^*$ & $+$ & $+$ & $+$ & $0$ & 250 & 900 \nl
2B911127 & \nodata & \nodata & \nodata & \nodata & 190 & 250 \nl
2B911202$^*$ & $0$ & $+$ & $0$ & \nodata & 350 & 1300 \nl
2B911209$^*$ & $+$ & $+$ & $+$ & $+$ & 270 & 400 \nl
2B920218$^*$ & \nodata & $+$ & \nodata & \nodata & 220 & 150 \nl
2B920311 & $+$ & $+$ & $+$ & $+$ & 440 & 1150 \nl
2B920406$^*$ & $+$ & $0$ & $+$ & $0$ & 240 & 300 \nl
2B920513$^*$ & $+$ & \nodata & $+$ & \nodata & 220 & 200 \nl
2B920517$^*$ & $+$ & \nodata & $+$ & \nodata & 230 & 150 \nl
2B920525$^*$ & $+$ & $+$ & $+$ & $+$ & 380 & 2075 \nl
2B920617 & \nodata & $+$ & \nodata & \nodata & 145 & 175 \nl
2B920622 & $+$ & $0$ & $+$ & \nodata & 440 & 1550 \nl
2B920627$^*$ & \nodata & \nodata & \nodata & \nodata & 210 & 325 \nl
2B920711 & $+$ & $0$ & \nodata & $0$ & 510 & 650 \nl
2B920720 & $+$ & $-$ & $-$ & $-$ & 270 & 525 \nl
2B920902$^*$ & \nodata & $+$ & $+$ & \nodata & 550 & 1500 \nl
2B921009 & $+$ & $0$ & \nodata & \nodata & 300 & 225 \nl
2B921123 & $+$ & $+$ & \nodata & \nodata & 250 & 350 \nl
2B921207 & $+$ & $+$ & $+$ & $+$ & 140 & 650 \nl
2B921209 & $+$ & $+$ & $+$ & $+$ & 180 & 375 \nl
2B930120 & \nodata & \nodata & \nodata & \nodata & 125 & 80 \nl
3B930405 & \nodata & $-$ & \nodata & \nodata & 280 & 300 \nl
3B930425 & \nodata & \nodata & \nodata & \nodata & 250 & 225 \nl
3B930506 & $+$ & $+$ & $0$ & \nodata & 1030 & 1750 \nl
3B930916$^*$ & $0$ & $0$ & $0$ & $0$ & 390 & 350 \nl
3B930922 & $+$ & $+$ & $+$ & $0$ & 110 & 130 \nl
9 3B931103 & $+$ & $+$ & $+$ & $+$ & 380 & 1300 \nl
3B931126 & $+$ & $+$ & $+$ & $+$ & 240 & 300 \nl
3B931204 & $+$ & $+$ & $+$ & $0$ & 380 & 775 \nl

\tablenotetext{*}{$\beta=-2.01$ throughout the burst.}
\tablenotetext{\dagger}{Observed by COMPTEL (\cite{han94}).}
\tablenotetext{~}{Burst Characteristics:
\begin{verse}
1 --- \ep-intensity association.\\
2 --- Overall softening within a burst.\\
3 --- Softening within pulse structures.\\
4 --- Later spikes softer than earlier spikes.\\
\end{verse}}
\tablenotetext{~}{The symbols denote:
\begin{verse}
$+$ --- Trait observed.\\
$-$ --- Opposite behavior observed.\\
$0$ --- Neither trait nor opposite observed.\\
$\ldots$ --- Could not be determined.\\
\end{verse}}

\end{planotable}

\clearpage

\begin{planotable}{rrl}
\tablewidth{350pt}
\tablecaption{Summary of Classification 37 Bursts}
\tablehead{
\colhead{\# Observed} & \colhead{\# Possible} &
\colhead{Property}}
\startdata
23 & 25 & \ep -- intensity association \nl
\noalign{\vskip 7pt}
20 & 30 & \ep ~softens over whole burst \nl
2  & 30 & \ep ~hardens over whole burst \nl
\noalign{\vskip 7pt}
18 & 22 & \ep ~softens within intensity spikes \nl
1  & 22 & \ep ~hardens within intensity spikes \nl
\noalign{\vskip 7pt}
9  & 17 & Later spikes softer than earlier ones \nl
1  & 17 & Later spikes harder than earlier ones \nl
\end{planotable}

\clearpage

\begin{planotable}{lcc}
\tablewidth{250pt}
\tablecaption{Unbiased Burst Separation}
\tablehead{
\colhead{} &
\colhead{\dep$ < 400$ keV} &
\colhead{\dep$ \ge 400$ keV}}
\startdata
$t_d \ne 0$ & 10 & 7 \nl
$t_d = 0$ & 8 & 12 \nl

\tablecomments{This table shows little evidence that bursts in which
emission is delayed evolve differently than those which emit promptly. However,
although the criteria used to divide the sample is unbiased, it may not be
appropriate since most bursts emit very soon after the trigger.}

\end{planotable}

\clearpage

\begin{planotable}{rcccc}
\tablewidth{300pt}
\tablecaption{Bursts Divided According to Emission Time}
\tablehead{
{$t_d\over\Delta T$} &
Number &
Min(\dep)  &
Med(\dep)  &
Max(\dep) \\
\colhead{} & \colhead{} & \colhead{(keV)} & \colhead{(keV)}& \colhead{(keV)}}
\startdata
$>0.25$     & 10 & 80 & 300 & 950 \nl
$\leq 0.25$ & 27 & 120 & 525 & 2075 \nl
\tableline
$>0.5$      & 8 & 80 & 300 & 950 \nl
$\leq 0.5$  & 29 & 120 & 525 & 2075 \nl
\tableline
$>0.9$      & 7 & 80 & 300 & 950 \nl
$\leq 0.9$  & 30 & 120 & 525 & 2075 \nl
\tableline
$>1$        & 5 & 80 & 300 & 950 \nl
$\leq 1$    & 32 & 120 & 440 & 2075 \nl

\tablecomments{This table summarizes the differences in evolution of \ep ~for
bursts grouped according to the time at which they emit.}

\end{planotable}

\begin{figure}
\caption{\ep~as determined from spectra at the finest available time resolution
(crosses) and from pairs of spectra averaged together (diamonds). The data
are from burst 2B910807 and the size of the symbols represents the
time intervals over which spectra were accumulated and the error in \ep.}
\end{figure}

\begin{figure}
\caption{\ep~ as determined from fits to the same spectra with different
binnings for burst 3B930916. The crosses represent values from the
unrebinned data while the diamonds correspond to the values from spectra
which have been rebinned into eight broad energy bins.}
\end{figure}

\begin{figure}
\caption{\ep~as measured by detectors at different gains for burst 2B910503.
The high gain detector (crosses) was fit from 25.7-2977 keV and the low gain
detector (diamonds) covers 308-26,214 keV. The high gain detector has better
time resolution since it had a stronger signal.}
\end{figure}

\begin{figure}
\caption{\ep~as determined from two detectors viewing the burst
2B921207. The high gain detector (crosses) covered the energy range from
20.2-1283 keV and the low gain detector (diamonds) ranged from 43.1-3300 keV.}
\end{figure}

\begin{figure}
\caption{The evolution of \ep~and the count rate for all bursts in the
sample. The histogram is the count rate for individual SHERB spectra and the
diamonds are \ep. The height of the diamonds represent 1-$\sigma$ error bars
and their width indicates the accumulation time of the SHERB spectra averaged
together to achieve the S/N criterion outlined in $\S$3. Note: 3B931204 extends
over two frames.}
\end{figure}

\begin{figure}
\caption{A comparison of the peak in \ep~between intensity pulses within the
same burst. The diagonal line indicates equality. Harder early pulses lie
below this line and harder late pulses above it. The bursts used are those
for which a $+$, $-$, or $0$ answer could be determined for column~4 in
Table~2. Several points line up vertically and horizontally since early pulses
were individually compared to all those which followed.}
\end{figure}

\begin{figure}
\caption{The range in \ep ~for a given emission time for all bursts in the
sample. The vertical bars give the range in \ep ~(max-min) and the horizontal
bars show the period during which significant emission occurred relative to
trigger. The crosses mark median values.}
\end{figure}

\end{document}